\documentclass[12pt]{iopart}
\usepackage{iopams,graphicx} 

\begin{document}

\title[Small Angle Approximation for non parallel plate capacitors]{Small Angle Approximation for non parallel plate capacitors with Applications in Experimental Gravitation}

\author{B R Patla }

\address{Smithsonian Astrophysical Observatory,\\
 Harvard-Smithsonian Center for Astrophysics,\\
60 Garden St, Cambridge, MA 02138, USA 
}
\ead{bpatla@cfa.harvard.edu}
\begin{abstract}
An approximate analytical formula for the capacitance of a non-parallel plate capacitor with small values of inclination angles and distance separations  of the plates is presented. Most applications involving position sensing or stabilizing the proof masses of precision gravity experiments~(for example, testing the equivalence principle) often use a parallel plate approximation for modeling the capacitance, which in turn, is used for estimating the noise. The analytical approximation presented here is based on the more general, but hard to implement, formalism for small angles given in Xiang (2005). Our approximation is accurate and reliable in the small-angle limit because it does not involve elliptic integrals of the first kind that has an indeterminate value at unity. Our formalism is very useful for computing the forces acting on the proof masses simply by taking the derivative of the capacitance. 

Effects of varying the inclination angle, plate dimension, and distance separation on the value of the capacitance per unit length are analyzed in detail.
We use the formula derived in this paper to compute the acceleration of the proof mass due to the presence of sensing electrodes (used for positioning the proof masses) with an assumed tilt of $\sim 10^{-6}$\;rad as applicable to SR-POEM: an experiment aiming to test the weak equivalence principle -- a fundamental postulate of general relativity~\cite{rdra,rdrb}.
\end{abstract}


\maketitle

\section{Introduction}
Capacitive sensing is one of the most commonly used techniques in science and engineering for position measurements ranging from a few millimeters up to slightly less than a nanometer. In experimental physics, gravity experiments in particular, often a parallel plate approximation is used for estimating the noise due to the presence of positioning electrodes near the proof masses, or as in some cases, for deducing the noise component---most often being the primary source of error---in distance measurements using capacitance gauges. Examples include gravitational wave detectors~\cite{speakea,speakeb}, precision experiments like SR-POEM that intends to test the weak equivalence principle (WEP)~\cite{rdra,rdrb}, torsion balance experiments \cite{ega}, etc.

More recently, the effects of assuming a non-parallel plate capacitor model for the switches used  in micro electro mechanical systems (MEEMS)~\cite{brown} and  differential micro accelerometers used in micromechanical machining~\cite{tay} have both concluded that using the simplified parallel plate model underestimates the capacitance by a few percent. This tiny fraction, perhaps negligible for ordinary applications, is nevertheless a factor that cannot be overlooked when the intended measurements demand  sub-picometer level precision requirements. In this paper, we will derive a formula that is very useful for computing capacitance with an acceptable margin of error.

In the case of WEP test, before the actual measurement of the precise positions of the proof masses  in free-fall commences, the proof masses have to be positioned accurately using electrostatic force within the housing. If the proof masses are let off non-parallel with respect to the sensing and forcing electrodes after positioning, the resulting configuration will introduce a component of force (on the proof masses)  along the direction of the WEP measurement, which will be slightly more than the corresponding configuration mimicking a parallel plate configuration. A rough estimate of the force can be obtained by taking the derivative for the capacitance of two parallel plates, $\epsilon_0 A/d$, where $A$ is the area of the plate, $d$ the distance separating the plates and $\epsilon_0$ is the vacuum permittivity. 

The general formula for non-parallel plates as given in~\cite{xiang} involves elliptic integrals that have to be evaluated numerically. For tiny distance separations and small angles that are of interest in experimental gravitation, the integrals evaluate to infinity even with reasonable numerical precision because of the very nature of elliptic integrals being singular at unity (and enormously large even around unity). This requires one  to impose cutoffs resulting in an overdetermination or underdetermination of the capacitance, depending on what stage of the numerical evaluation the cutoffs are imposed. In order to obtain reasonable estimates of the capacitances to first order approximation, a convenient formula for capacitance that establishes a functional relationship between angle and distance separating the plates is more useful. Moreover, the force between the plates is directly proportional to the derivative of the capacitance and the closed form approximate solution of the capacitance is useful for computing the force directly and more accurately.

In section~\ref{sec:2} we provide a brief and transparent overview of the results in~\cite{xiang}, whose main result we will refer to as semi-analytical from hereafter and for the rest of this paper because of the elliptic integrals involved in the final form that have to be evaluated numerically, and so it is not a closed form solution in the conventional sense. To make a clear distinction, the results derived in section~\ref{sec:3} will be dubbed as the small angle approximation, which is valid only for small values of inclination angles and separations (much smaller than the linear dimensions of the plates). The result of this section is compared with finite element analysis (FEA) model and also the semi-analytic formula. Section~\ref{sec:4} delineates the effects of the relative orientation of the plates using the semi-analytic formula. There is no difference in the value of capacitances to first order, using the formula for small angle approximation. In section~\ref{sec:5} we will use our small angle approximation formula to study the spurious WEP (acceleration) signal in SR-POEM , followed by a brief discussion in section~\ref{sec:6}.

\section{\label{sec:2} Capacitance of two non-parallel plates}

\begin{figure}[htb]
\begin{center}
\includegraphics[width=0.49\textwidth]{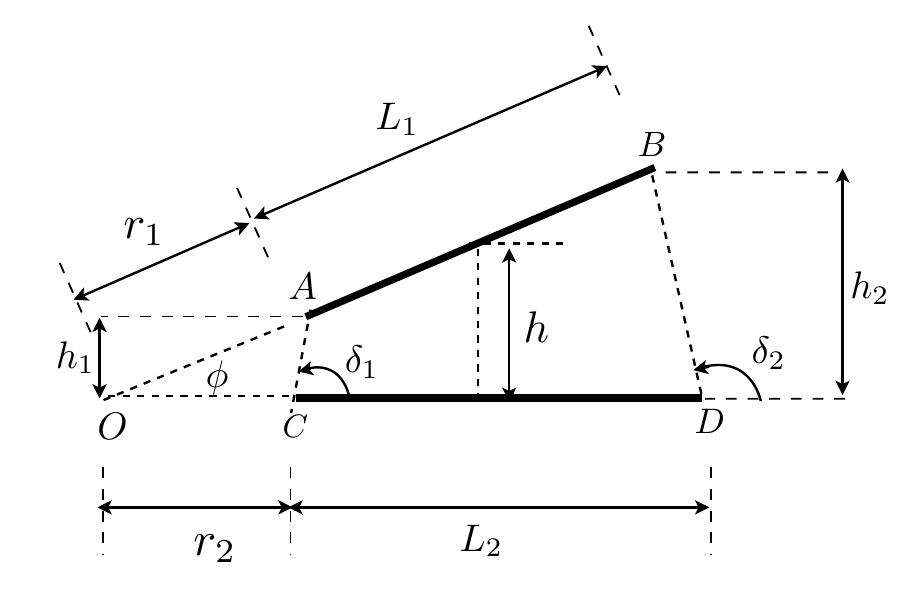}
\caption{Plates of the non parallel plate capacitor have widths   $L_1$and  $L_2$ respectively with their lengths extending infinitely along the  $z$ direction. The geometric centers of the plates coincide  along the horizontal axis ($\Delta x=0$, need not be true in general) and are separated by a distance $\Delta y=h$ along the $y$ axis. The extensions of the plates that intersect at a point $O$ are the lengths  $r_1$ and $r_2$ . The plates make an angular separation $\phi$  (measured counter clockwise from the $x$ axis). Points $A, B,  C$ and $D$  are the end points of the plates in the $xy$ plane.}
\label{fig:1}
\end{center}
\end{figure}

We briefly introduce the formalism developed in~\cite{xiang} to set the stage for arriving at an approximate solution for the capacitance per unit length in the next section.
Consider two infinite strips extending through the $z$ direction and subtending an angle $\phi$ between them,  see figure~\ref{fig:1}. 
The coordinates of the center of the upper plate are $(\Delta x, \Delta y)$ from the center of the bottom plate. The configuration is subjected to successive multiple coordinate transformations as given in Table~\ref{tab:1}. The first column of Table~\ref{tab:1} consists of identifiers for the two dimensional space representing the cross section of the plates. Initially the ends of the plates are mapped onto a the complex $z$-plane and followed by successive mappings.  The idea is to find a coordinate system in which the non-parallel plates of unequal widths are parallel to each other and also have equal widths.  The capacitance of such a configuration is to be identified with that of an ordinary parallel plate capacitor, albeit the linear dimensions corresponding to the aforesaid configuration being complex functions.

\Table{ \label{tab1} Set of successive coordinate transformations of the non-parallel plates  }

\br
&\centre{4}{Coordinates$^{\dagger}$}\\
\ns
& \crule{4}\\
Plane & A & B & C & D & Transformation  \\
\mr
$z$ &$r_1 e^{i \phi}$ & $(r_1+L_1) e^{i \phi}$& $r_2$ & $ r_2+L_2$ & \hspace{0.5in} --- \\
$t$ & ($\alpha$, 0) & $(-1, 0)$ &$( \beta, 0) $ & $(1, 0) $ & $t=c_1 z^{\pi/\phi}+c_0$$^{\star}$, $\phi\rightarrow\pi$\\
$\zeta$ & ($-\infty$, 0) & $(0, 0)$ & $(1/k^2, 0)$ & $(1, 0)$ & $\zeta=\frac{(1-\alpha)(1+t)}{2(t-\alpha)}$\\
$u$ & $i K'$ & 0 & $K+iK'$ &$K$ & Schwarz-Christoffel\\

\br
\end{tabular} 
\item[] $^{\dagger}$ End points  of the  cross-section of the upper and the lower plates,  figure 1.
\item[] $^{\star}$ Constants $c_1$ and $c_0$ are determined by imposing the constraint --- $x$ coordinates of $B$ and $D$ be $-1$ and $1$ respectively.
\end{indented}
\label{tab:1}
\end{table}

The initial configuration in the old complex $z$ plane is transformed on to the new $t$ plane such that the ends of the plates are fixed at $+1$ and $-1$ respectively. These  constraints determine the functional form of the constants to be
\begin{equation*}
c_1=\frac{2}{(L_1+r_1)^{\pi /\phi }+(L_2+r_2)^{\pi /\phi }} \quad  \rm{and} 
\end{equation*}
\begin{equation}
c_0=\frac{(L_1+r_1)^{\pi /\phi }-(L_2+r_2)^{\pi /\phi }}{(L_1+r_1)^{\pi /\phi }+(L_2+r_2)^{\pi /\phi }},
\end{equation}
where  $L_1$ and $L_2$  are the widths of the plates respectively.
The transformation into the $\zeta$ plane (see Table~\ref{tab:1}) yields coordinate distances that are appropriate for the Schwarz-Christoffel transformation~\cite{mandf} that is to be invoked next, which maps the interior of the plates on to the area of a rectangle. Solving the equation $\zeta k^2-1=0,$ yields
\begin{equation}
k=\sqrt{\frac{2(\beta -\alpha )}{(1-\alpha )(1+\beta )}},
\label{expr4k}
\end{equation}
where 
\begin{equation*}
\alpha=\frac{-2 r_1^{\pi /\phi }+ (L_1+r_1)^{\pi /\phi }-(L_2+r_2)^{\pi /\phi }}{(L_1+r_1)^{\pi /\phi }+(L_2+r_2)^{\pi /\phi }} \quad \rm{and}
\end{equation*}
\begin{equation}
\beta=\frac{2 r_2^{\pi /\phi }+ (L_1+r_1)^{\pi /\phi }-(L_2+r_2)^{\pi /\phi }}{(L_1+r_1)^{\pi /\phi }+(L_2+r_2)^{\pi /\phi }}.
\label{expr4agbg}
\end{equation}
The Shwarz-Christoffel transformation maps the upper half of the complex plane into the interior of a rectangle.
\begin{equation}
\frac{du}{ d\zeta}= A_1 (\zeta -0)^{\frac{\pi }{2\pi}-1}(\zeta -1)^{\frac{\pi }{2\pi }-1}\left(\zeta -\frac{1}{k^2}\right)^{\frac{\pi }{2\pi }-1},
\end{equation}

\begin{equation}
u =A_1 k \int_0^{\zeta } \frac{1}{\sqrt{\zeta (1-\zeta )\left(1-\zeta  k^2\right)}}  d\zeta  +A_0.
\end{equation}
Substituting $\zeta=\rho^2$ and imposing the constraints at points $B (A_0=0)$ and $D (A_1 k=1)$ respectively, we identify

\begin{equation}
u_x\equiv K(k)=\int_0^1 \frac{1}{\sqrt{\left(1-\rho ^2\right)\left(1-\rho ^2k^2\right)}}  d\rho \quad \rm{and}
\end{equation}

\begin{equation}
u_y \equiv iK'(k)=\int_1^{1/k} \frac{1}{\sqrt{\left(1-\rho ^2\right)\left(1-\rho ^2k^2\right)}}  d\rho 
\end{equation}
where $ K(k)$ is the complete elliptic integral of the first kind of modulus $k$ with $K(k')=K'(k)$ and $k'^2=1-k^2$. From equation~(\ref{expr4k}), 
\begin{equation}
k'=\sqrt{\frac{(1+\alpha )(1-\beta ) }{(1-\alpha )(1+\beta )}}.
\label{expr4k'}
\end{equation}
The capacitance per unit length, using the parallel plate formula, is 
\begin{equation}
C=C_{\rm{in}}+C_{\rm{out}}=\epsilon_0\left(\frac{K'(k_{\rm{in}})}{K(k_{\rm{in}})}+\frac{K'(k_{\rm{out}})}{K(k_{\rm{out}})}\right)
\label{inccap}
\end{equation}
where $k_{\rm{in}}$ and $k_{\rm{out}}$ represent the values of $k$ (equation~(\ref{expr4k})) for the interior and exterior of the plates corresponding to the angles $\phi$ and $2\pi-\phi$ in equation~(\ref{expr4agbg}).

\section{\label{sec:3} Approximate formula for small angles}
Values of the elliptic integrals in equation (\ref{inccap}) have to be evaluated numerically. For very small values of $\phi$ and $h/L$ (see figure~ \ref{fig:1}), even after allowing a reasonably high numerical precision,  we encounter indeterminate values for the integrals and so one has to impose cutoffs very close to unity. These numerical manipulations often introduce errors that result in over or under determination of the capacitances based on the numerical algorithms used. In order to compute the force as a function of distance separating the plates and the included angle between them, we have to numerically evaluate the value of capacitances for a range of angles and separations followed by numerical differentiation. 
We will use a simplified model for the capacitor plates by demanding their lengths be equal, $L_1\approx L_2\sim L$.
From  $\triangle \;AOC$,
\begin{equation}
r_1\approx \frac{h_1}{\sin\phi}, \quad r_2 \approx \frac{h_1 \sin(\delta_1-\phi)}{\sin\phi}
\label{subs1}
\end{equation}
\begin{equation}
r_1+L_1 \approx \frac{h_2}{\sin\phi}, \quad r_2+L_2 \approx \frac{h_2\sin(\delta_2-\phi)}{\sin\phi}
\label{subs2}
\end{equation}
\begin{equation}
h_1=h-\frac{L}{2}\phi, \quad h_2=h+\frac{L}{2}\phi
\label{subs3}
\end{equation}
 From equations~(\ref{expr4k'}) and (\ref{expr4agbg}), 
\begin{equation}
{k'}_{\rm{in}}^{2}= \left(\frac{(r_1+L_1)^{\pi/\phi}-r_1^{\pi/\phi}}{(r_1+L_1)^{\pi/\phi}+r_2^{\pi/\phi}}\right)\left(\frac{(r_2+L_2)^{\pi/\phi}-r_2^{\pi/\phi}}{(r_2+L_2)^{\pi/\phi}+r_1^{\pi/\phi}}\right),
\label{expr4k'2}
\end{equation}
where $\delta_1=\pi/2-\phi, \delta_2=\pi/2+\phi$ and $L_1=L_2\equiv L$. Using equations~(\ref{subs1}) and (\ref{subs2}) in equation~(\ref{expr4k'2}) yields 
\begin{equation}
{k'}_{\rm{in}}^{2}= \tanh\left(\frac{\pi L}{2 h}\right) \tanh\left(\pi\phi+\frac{\pi L}{2h}\right).
\label{expr4k'2f}
\end{equation}
equation~(\ref{expr4k'2f}) is valid when $h/L>\phi$ and $\phi\ll 1$, necessary conditions for small angle approximation. Note that, the usual condition for parallel plate capacitor, $L/h\gg1$($\sim 10$ is sufficient in most cases), must also be satisfied independently.  Letting 
\begin{equation}
\frac{\pi L}{2h}=a \quad {\rm and} \quad \pi\phi+\frac{\pi L}{2h}=b
\label{aandb}
\end{equation}
 where $a,b\gg1$, yields
\begin{equation}
\sqrt{k'} \sim 1-\frac{e^{-2 a}}{2}\left(1+e^ {-2\pi \phi}\right).
\end{equation}
The relationship between the elliptic integrals and their complementary modulus may be expressed in terms of the nome
\begin{equation}
\frac{K'(k)}{K(k)}=-\frac{1}{\pi}\log q,
\end{equation}
where $q$ is the nome which may also be expressed as a functional of {\it theta} functions, denoted by $\lambda$ (see appendix for a derivation).
\begin{equation}
 \lambda_{\rm{in}}=\frac{1-\sqrt{k'}}{2(1+\sqrt{k'})} \sim e^{-2a}\left(1-\pi\phi\right)\left(4- 2e^{-2a}\left(1-\pi\phi\right)\right)^{-1},
 \label{lambda_in}
  \end{equation}
ignoring the second and higher order terms in the expansion of the exponential term. 

In order to evaluate the exterior capacitance for the configuration in  figure~\ref{fig:1} by using equation~(\ref{inccap}), we need an expression for  $k'$ outside. Replacing $\phi\rightarrow 2\pi-\phi$ in equation~(\ref{expr4k'2}) one obtains ${k'}_{\rm{out}}^{2} = {k'}_{\rm{in}}^{2}|_{\phi\rightarrow 2\pi-\phi}\Rightarrow  {k'}_{\rm{out}} \sim L\phi/4h $, yielding
\begin{equation}
\lambda_{\rm{out}}=\frac{1}{2}\left(1-\sqrt{\frac{L\phi}{4h}}\right)\left(1+\sqrt{\frac{L\phi}{4h}}\right)^{-1},
\label{lambda_out}
\end{equation}
where we replaced $k'_{\rm in}$ with  $k'_{\rm out}$ in equation~(\ref{lambda_in}).
The capacitance per unit length, to first order approximation in $\phi$ is 
\begin{equation}
C^{(1)}=-\frac{\epsilon_0}{\pi}\left(\log \lambda_{\rm{in}}+\log \lambda_{\rm{out}}\right).
\label{cap1st}
\end{equation}
Including higher order terms in $\lambda$ and using equations~(\ref{lambda_in}) and (\ref{lambda_out}) in equation~(\ref{cap1st}), we obtain an expression for the capacitance
\begin{eqnarray}\nonumber
C =\frac{\epsilon_0 L}{h}-\frac{\epsilon_0}{\pi}\left[\log\left(1-\pi\phi \right)+\log\left(1-\sqrt{\frac{L\phi}{4h}}\right)-\log\left(1+\sqrt{\frac{L\phi}{4h}}\right)\right] \\
\nonumber
+ \frac{\epsilon_0}{\pi}\left[ \log 4 \left(2-(1-\pi\phi)\exp\left(-\frac{\pi L}{h}\right)\right)\right] \\
 -\frac{\epsilon_0}{\pi} \left[ \log(1+2\lambda_{\rm{in}}^4+\cdots) +\log(1+2\lambda_{\rm{out}}^4+\cdots)\right].
 \label{capfinal}
\end{eqnarray}
The two terms in the first line of equation~(\ref{capfinal}) comprises the equivalent of the formula for the parallel plate capacitor (in the limit of $\phi\rightarrow 0$) and the last two terms represent the first order fringe capacitance and the rest of the higher order terms. Series expansion of the terms in the first line of equation(\ref{capfinal}) and retaining only first and second order terms in $\phi$ yields
\begin{equation}
C = \frac{\epsilon_0 L}{h}+\frac{\epsilon_0}{\pi}\left[\left(1+\frac{L}{4h}\right)\phi +\frac{\phi^2}{2} \right].
\end{equation}

\begin{figure}[b]
\begin{center}
\includegraphics[width=0.6\textwidth]{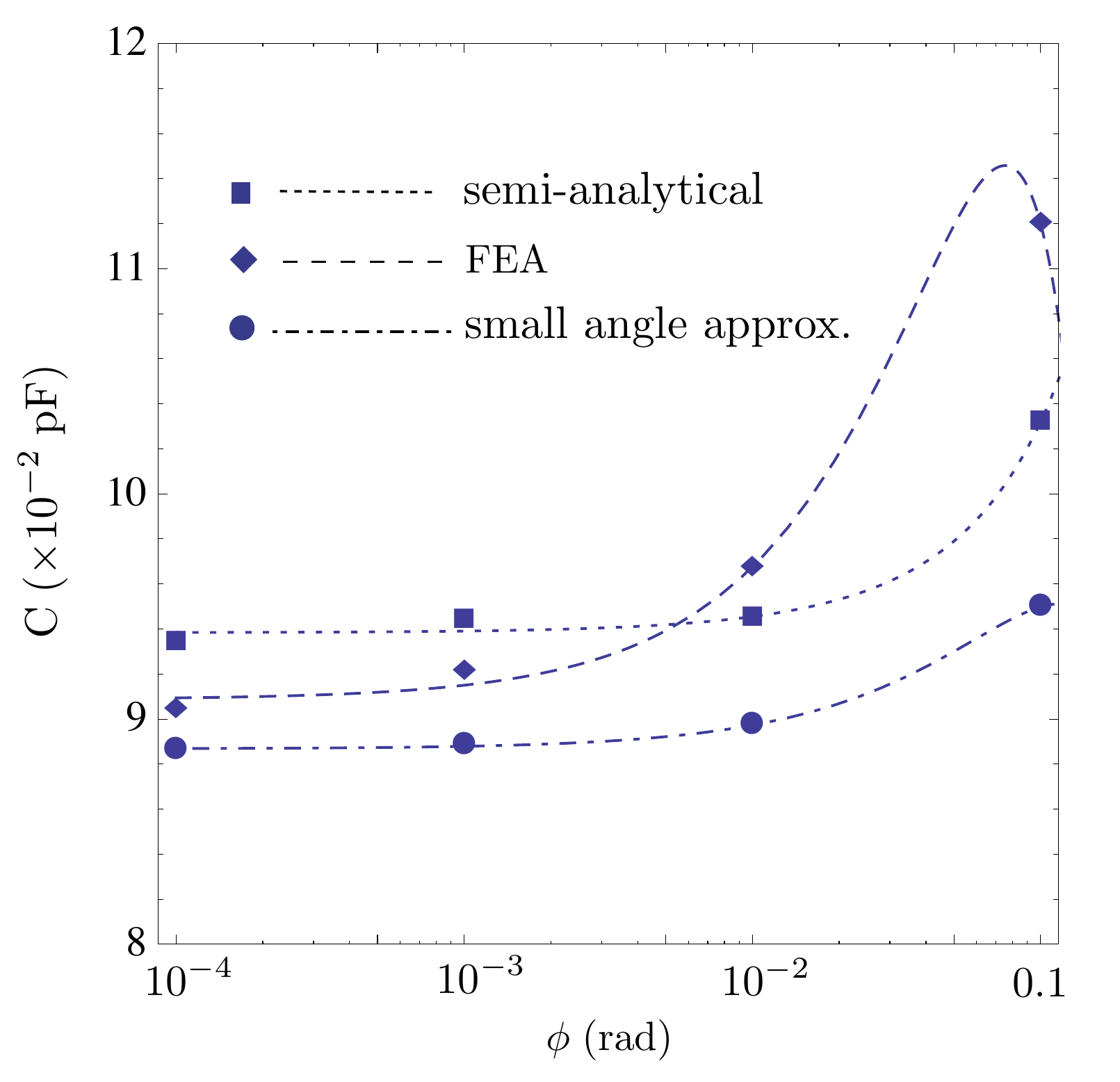}
\caption{Capacitance corresponding to a pair of plates each with sides 1\;mm separated by a distance of 0.1\;mm for angles $10^{-4}, 10^{-3}, 10^{-2} $ and 0.1\;rad.
The data points are obtained by using three different methods: 1) Semi-analytical from~\cite{xiang} 2) Finite Element Analysis (FEA)  3) Small angle approximation. The upper plate is tilted about its center with the bottom plate fixed. The curves connecting the data sets are  quadratic best fits corresponding to the three cases. The error in the FEA is less than 5\% for the maximum values of $\phi$ in the figure. Fringing capacitance is not included in this plot. The data points corresponding to $\phi=10^{-3}$ and $ 10^{-2}$~radian in evaluating the values for capacitances using the semi-analytical case have large uncertainties~($\sim5\%$) as a result of the cutoffs imposed in the numerical evaluation of elliptic integrals.} 
\label{fig:2}
\end{center}
\end{figure}

The main result of this section(equation(\ref{capfinal})) is compared with numerical  simulations (FEA)  and ~\cite{xiang}, and are plotted in figure\;\ref{fig:2}. 
For the FEA, when $\phi=0$, the value of the capacitance match with the standard parallel plate capacitor formula to within 2\%. As the tilt of the upper plate is increased, the error in the estimated value of the capacitance increases due to the leaking of the fringe capacitance into the bounding box. The small angle approximation matches with the FEA more than it does for the semi-analytical formula due to over estimation of the capacitance introduced as a result of cutoffs imposed on the numerical estimation of the elliptic integrals. 

The indeterminacy of the elliptic integrals  near unity is a vexing problem as far as numerical evaluation of the capacitance is concerned for certain plate geometries. For example, if one were to use python, the accuracy for floating point variables is limited to $1-10^{-n}$, where $n=308$. In order to avoid using cutoffs we often need an accuracy corresponding to values of $n=500$ or higher. Furthermore, due to the involvement of product terms, this problem gets worse because each individual term places even more stringent constraints on the value of $n$.

Note that equation~(\ref{inccap}) contain ratios of elliptic integrals and that is precisely the reason why the value of the capacitance is over determined for some combinations of plate geometries (distance of separation, angle, widths) and under determined for others. However, for values of $\phi\sim10^{-2}$ and above, the small angle approximation is inaccurate and the semi-analytic approach is more reliable as the elliptic integrals now yield finite values. Therefore, the approximate solution for small angles is a rather convenient way to estimate the capacitance and the force for very small angles that we often encounter in precision gravity experiments. For the small angle approximation, the capacitance is now a function of distance separating the plates and inclination angle.

\section{\label{sec:4} Capacitance as a function of  rotation angle: Two simple cases}
Although for small angles and separations the choice of the rotation axis of the upper plate (figure~\ref{fig:1}) has no effect on the capacitance to first order in $\phi$, it matters for large values of inclination angles. 
This is due to the difference in the values  of the effective separation between the plates  corresponding to the choice of rotation axes. The semi-analytic approach outlined in~\cite{xiang} is used to investigate two cases corresponding to two different axes of rotation, for inclination angles that are greater than $10^{-2}$~rad.

 We consider two simple cases wherein the axes of rotations of the upper plate are 1) a line passing through the middle (along the $z$ direction in figure~\ref{fig:1}) of the upper plate and 2) an edge of the upper plate directly above a corresponding edge of the lower plate. We will study the variation of capacitance with angle, plate separation, and relative dimensions of the plates.

\subsection{Rotation about the centerline of the upper plate}
We introduce a coordinate system as described in figure~\ref{fig:1}

\begin{equation}
r_1=\frac{\Delta y}{\sin\phi}-\frac{L_1}{2}\ ,  \quad  r_{1\rm{min}}=\sqrt{\left(\frac{L_2}{2}+\Delta x\right)^{2}+{\Delta y}^2}-\frac{L_1}{2},
\label{midpiv1}
\end{equation}
\begin{equation}
r_2=\frac{\Delta y}{\tan\phi}-\Delta x-\frac{L_2}{2} \ ,  \quad  \phi_{\rm{max}}=\sin^{-1}\left( \frac{\Delta y}{\frac{L_1}{2}+ r_{1\rm{min}}}\right),
\label{midpiv2}
\end{equation}
where $ r_{1\rm{min}}$ is the minimum distance of the nearest edge of the upper plate from the point of intersection of the plates and $\phi_{\rm{max}}$ is the maximum value allowed for the rotation angle, with all other notations retaining the same definitions  as that of in the caption of  figure~\ref{fig:1} of section~\ref{sec:2}. Using equations~(\ref{midpiv1}) and (\ref{midpiv2}) in equation~(\ref{inccap}) we proceed to study the capacitance as a function of included angle and relative linear dimensions of the plates for  distance separations of 5~mm, 7.5~mm and 1~cm respectively.
The chosen separations are the typical values encountered in SR-POEM -- the spacing between the proof mass and housing -- discussed at length in section~\ref{sec:5}. Any values for separations that are less than 5~mm, for angles $10^{-2}$~rad or less, requires the imposition of cutoffs for evaluating the elliptic integrals.

\begin{figure}[htb]
\begin{center}
\includegraphics[width=0.9\textwidth]{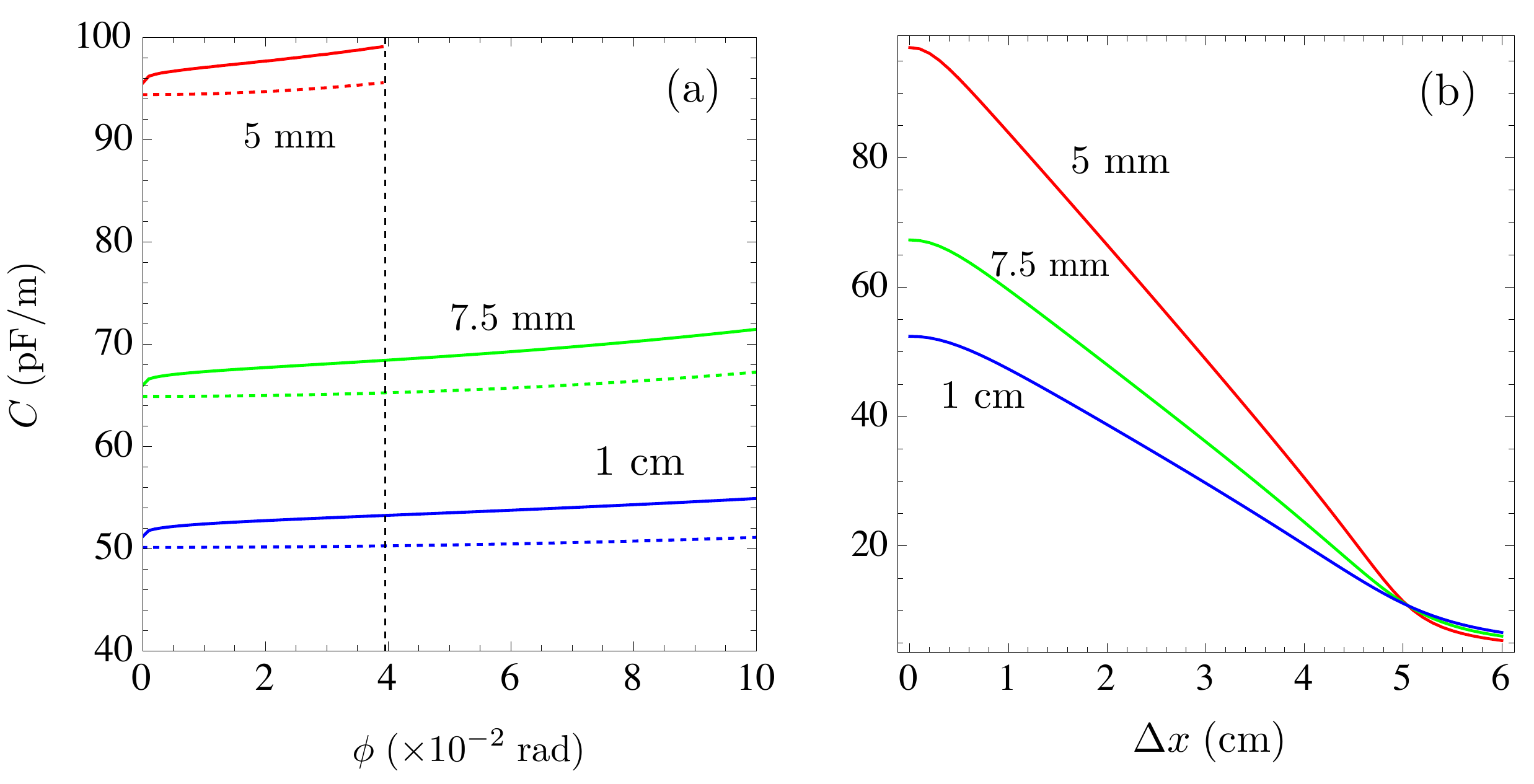}
\caption{Variation of capacitance per unit length for a pair of plates, each with width 5\; cm for distance separations corresponding to 5\;mm, 7.5\;mm and 1\; cm respectively. 
(a) As a function of included angle $\phi$. For 5~mm case, the plates touch when the value of $\phi$ exceeds 40~mrad (vertical dashed line). 
(b) As a function of plate offset $\Delta x$. The included angle is fixed at $10^{-2}$\; rad.
}
\label{fig:3}
\end{center}
\end{figure}

For plates of equal widths, the effective separation decreases for the left half of the configuration and thereby increasing the capacitance. The total capacitance is  the sum of the capacitances as if they are in parallel yielding a smaller effective separation (product of the distance separations in the formula for two capacitances in parallel) resulting in a higher capacitance as the angle is increased.  As the offset along the $x$ direction ($\Delta x$) is increased, the plates overlap less if their linear dimensions are held constant, thereby reducing the capacitance. As the included angle is increased the effective  separation of the left half increases, contributing even more to the total capacitance. These results are plotted in figure~\ref{fig:3}
\subsection{\label{subsec:1} Rotation about an edge of the upper plate}
\begin{figure}[htb]
\begin{center}
\includegraphics[width=0.9\textwidth]{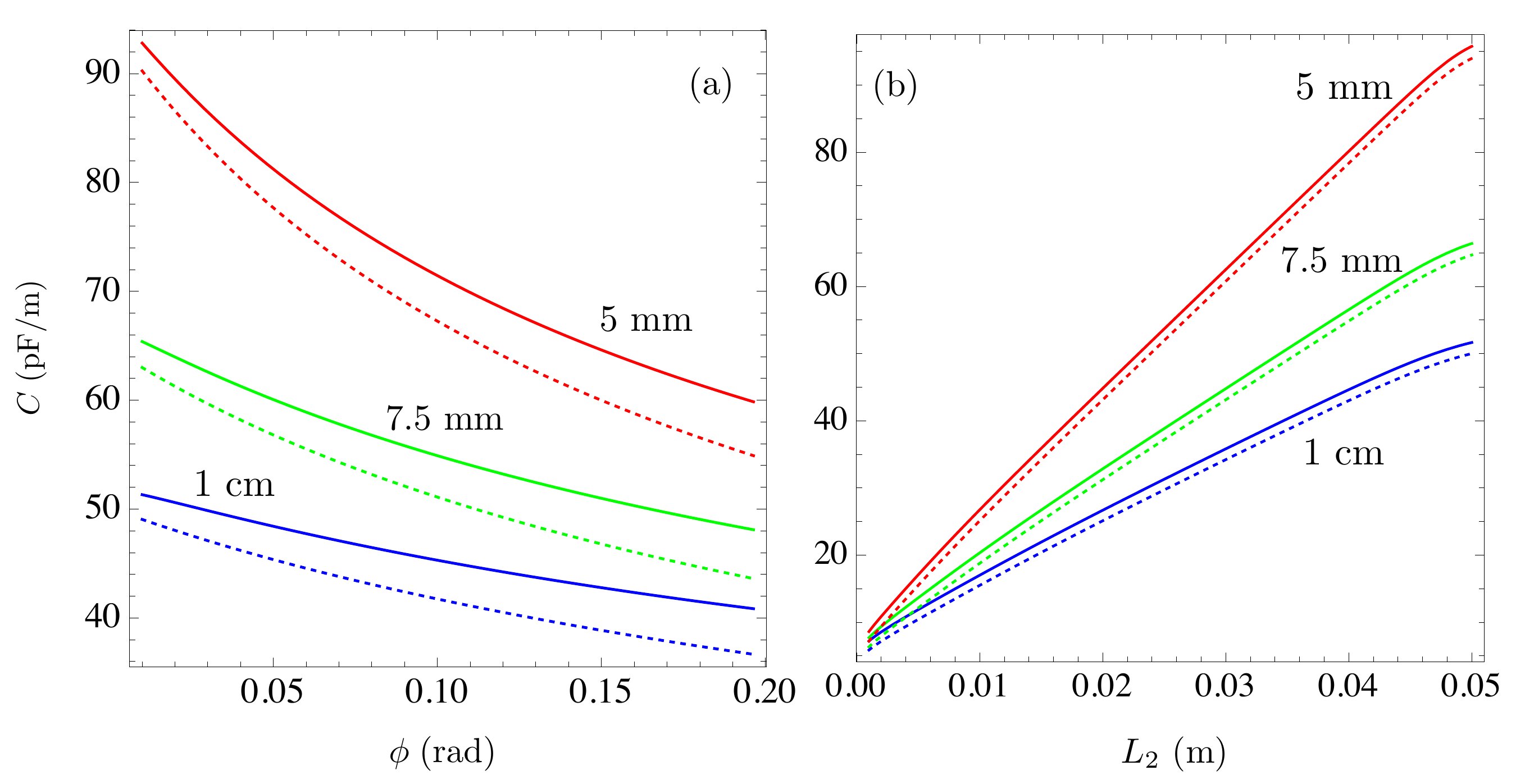}
\caption{Variation of capacitance per unit length for distance separations corresponding to 5\;mm, 7.5\;mm and 1\; cm respectively. The dashed line includes the fringing capacitance, which is less than 10\% of the total value. (a) The widths of the plates are fixed at 5\; cm each.
 (b)~The upper plate  is set at an angle of  $10^{-2}$\; rad with respect to the lower plate and $L_1=5$~cm.
}
\label{fig:4}
\end{center}
\end{figure}

We introduce a coordinate system by redefining 
\begin{equation}
r_1=\frac{\Delta y}{\sin\phi} \quad {\rm and}  \quad r_2= \frac{\Delta  y}{\tan\phi}.
\end{equation}
where $\Delta y\equiv h$.
The plate separation is now the distance separating  the aligned edges of the two plates of equal widths and $\phi$ being the inclination of the upper plate with respect to the horizontal lower plate. The variation of capacitance as a function of $\phi$ and relative linear dimensions are shown in figure~\ref{fig:4}. 
The capacitance decreases with $\phi$ as opposed to the case when the axis of rotation was directly above the center of the lower plate( section~\ref{subsec:1}). This is because the effective distance is now directly proportional to $\phi$. Unlike in the previous case, as the value of $\phi$ increases, the effective distance of separation also increases, $\Delta d\propto L_2 \phi$, where $\Delta d$ is the increment for the distance separating the plates. Therefore, the decrease in capacitance as a function of angle is primarily due to the increase in effective separation.  The functional dependence of capacitance on the linear dimensions ($L_1$ and $L_2$) of the plates exhibit  a similar pattern to that of  the configuration described  in section~\ref{subsec:1}, see figure~\ref{fig:4}.

\section{\label{sec:5} Test of the weak equivalence principle: A case study}

Testing of the equivalence principle as in SR-POEM amounts to the determination of $\eta$ parameter (measure of relative acceleration of two test masses of different materials) to an accuracy of  $10^{-17}$m/s$^2$ ~\cite{rdra}. A  laser gauge monitors the positions of the test masses, as the proof masses inside the vacuum chamber become part of a high finesse Fabry-Perot cavity. Before the measurements are made the test masses are properly aligned electrostatically by applying a drive voltage. When the actual distance measurements are performed, the capacitive sensing is turned off. However, the surface potential difference between the test mass and the sense electrodes that are used to align the test masses will generate a force along the measurement direction and produces a spurious acceleration signal. Many factors are responsible for generating a surface potential difference; patch effect being the most important one and that could be as high as a few milli volts~\cite{speakec,pollock}.

\begin{figure}[htb]
\begin{center}
\includegraphics[width=0.6\textwidth]{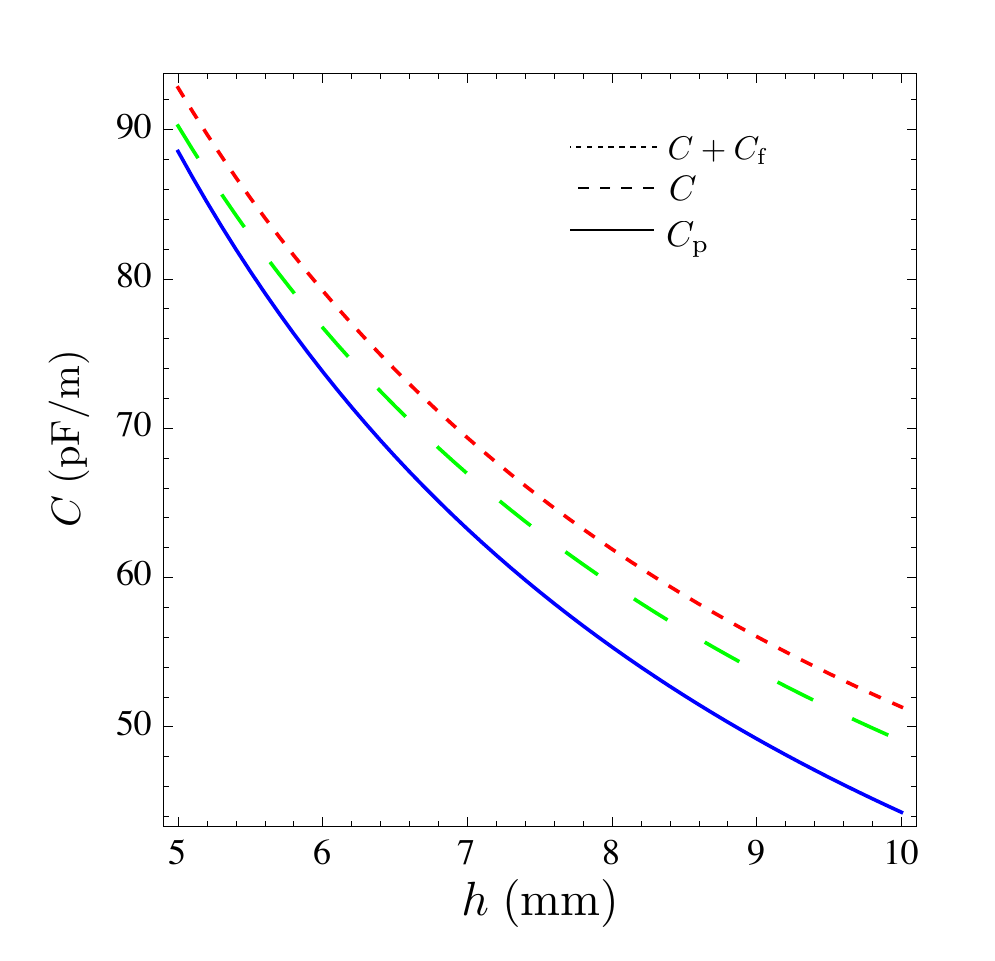}
\caption{Variation of capacitance per unit length as a function of distance. $C_f$ is the fringing capacitance, $C_P$ corresponds to the well-known formula for the parallel plate
case, $\epsilon_0 L/h$. Assumed plate dimensions are 5 cm with an inclination angle of $10^{-6}$~rad and a potential difference of $10^{-3}$~V between the plates.}
\label{fig:5}
\end{center}
\end{figure}

Even if the proof masses are properly positioned before the measurement, i.e. to say the sides of the proof masses are exactly parallel to the walls of the housing, there will still be a force acting on the proof mass in the direction perpendicular to the direction of the position measurement, and hence is of little concern to us. This is because a side of the proof mass together with a side of the housing will constitute the equivalent of  a parallel plate capacitor.
However, imperfections in machining  or the rotational inertia (as soon as the drive voltage is turned off) of the proof mass will cause a small tilt of the proof mass with respect to the housing walls. We will assume a maximum value of  $10^{-6}$~rad for this misalignment (tilt). A small component of the force  due to this tilt will result as an acceleration along the WEP measurement axis.
The force acting on the test mass is 
\begin{equation}
F=-\frac{1}{2} V^2 \nabla C(\phi,h)
\label{force}
\end{equation}
where $C(\phi,h)$ is the capacitance and $V$ is the (surface) potential difference across the plates. The resultant force acting on the test mass, for a relative inclination angle of $\phi$ with any given side of the housing, is 

\begin{equation}
F_s =\left(\frac{\pi}{1-\pi\phi}-\frac{s}{\pi\sqrt{\phi}(s^2-1)}\right)V^2\sin\phi
\label{sides}
\end{equation}
where $s=(l/4h)^{1/2}$, $l$ being the linear dimensions of the test masses and $h$ is the distance separating the test mass and the sensing electrode.
If there is a sensing  electrode on the bottom (although it is not a good idea to have one), the force is 
\begin{equation}
F_b=2s^3\left(4 s -\frac{\sqrt{\phi}}{\pi(s^2 -1)}\right)V^2 +\left(\frac{\pi}{1-\pi\phi}-\frac{s}{\pi\sqrt{\phi}(s^2-1)}\right)V^2\cos\phi .
\label{bottom}
\end{equation}
In deriving the above results we have neglected the fringing capacitance in equation~(\ref{capfinal}). The variation of the capacitance (with and without fringe effects) using small angle approximation derived in section~\ref{sec:3} is compared with the parallel plate case in figure~\ref{fig:5}.

For SR-POEM, estimate the force on the test masses surrounded by  sensing electrodes of  sides 5\; cm and separated by a nominal distance  5\;mm  from the test mass, allowing for a tilt of $\phi=10^{-6}$\;rad and a potential difference of a milli volt. 
\begin{enumerate}
\item{
 A plate facing a side of the test mass with the measurement direction perpendicular to the sides of the test mass will produce a force $4.48\times 10^{-21}$\;N}
\item{A bottom plate of similar dimensions yields a higher force $4.64\times 10^{-15}$\;N. This configuration is usually avoided.}
\item{ A parallel plate capacitor formula yields a force $8.85\times 10^{-16}$\;N}
\end{enumerate}

\section{\label{sec:6} Discussion}
We have studied the effects of non-parallel plate capacitors in the context of experimental gravitation where the angles are very small. An approximate formula to compute the capacitance in this regime is thoroughly analyzed along with the semi-analytic approach given in~\cite{xiang} and finite element methods. For very small angles the small angle approximation matches well with the finite element simulations (with maximum 2\% error in the appropriate angular regime). Around and after $\phi\sim10^{-2}$~rad the semi-analytical formula is a closer match.

In conclusion, we have shown that the small angle approximation for estimating the capacitance of non-parallel plate capacitors is a very good alternative to computationally expensive numerical methods and semi-analytical approaches. For applications in experimental gravitation it is easier to compute the forces, acceleration and also estimate other noise sources due to positioning of test masses  using our approximate solution. Our analysis successfully demonstrates that for an assumed surface potential difference of $10^{-3}$~V or less SR-POEM will easily meet it's mission goal of $10^{-17}$m/s$^2$.

\appendix
\section{Relationship between elliptic- and theta-functions}
The nome $q$ is defined as 
\begin{equation}
q=\exp(-\pi K'/K).
\end{equation}
The elliptic functions may be represented in terms of the {\it theta} functions
\begin{eqnarray}
\nonumber
\vartheta_1(z,q)&=&\sum_{n=-\infty}^\infty(-1)^{n-\frac{1}{2}}q^{(n+\frac{1}{2})^2} e^{(2n+1)i z}\\
\nonumber
\vartheta_2(z,q)&=&\sum_{n=-\infty}^\infty q^{(n+\frac{1}{2})^2} e^{(2n+1)i z}\\
\label{theta}
\vartheta_3(z,q)&=&\sum_{n=-\infty}^\infty q^{n^2} e^{2ni z}\\
\nonumber
\vartheta_4(z,q)&=&\sum_{n=-\infty}^\infty(-1)^{n}q^{n^2} e^{2ni z}\\
\nonumber
\vartheta_j(z+\pi,q)&=&\mp \vartheta_j(z,q)  \; \rm{negative\;sign\; for}\; j=1,2\\
\nonumber
\vartheta_j(z+\pi\gamma,q)&=&\mp N \vartheta_j(z,q)  \; \rm{negative\;sign\;for}\; j=1,4\\
\nonumber
\end{eqnarray}
where $N=q^{-1}e^{-2iz}$. A convenient way to obtain the ratio $K'/K$, is by taking the logarithm of the nome $q$.  Using the {\it theta} functions above, it is possible to arrive at a differential equation, the solution of which is an elliptic integral~\cite{mandf}. The form of the {\it modulus} is 
\begin{equation}
k=\vartheta_2(0,q)^2/\vartheta_3(0,q)^2\equiv\theta_2^2/\theta_3^2,
\label{ktheta}
\end{equation}
where we have adopted a new notation for the {\it theta} function corresponding to  $z=0$. Below we write down the series expansions using equation(\ref{theta}) for $z=0$
\begin{eqnarray}
\nonumber
\theta_2=&2 q^{1/4}+2 q^{9/4}+2 q^{25/4}+2 q^{49/4}+2 q^{81/4}+2 q^{121/4}+\cdots \\
\label{th-exp}
\theta_3=&1+2q+2q^4+2q^9+2q^{16}+2q^{25}+2q^{36}+\cdots \\
\nonumber
\theta_4=&1-2q+2q^4-2q^9+2q^{16}-2q^{25}+2q^{36}+\cdots \\
\nonumber
\end{eqnarray}
noting that $\theta_1=0$. Also, since $\theta_2^4+\theta_4^4=\theta_3^4$, using equation (\ref{ktheta})  we have $k'\equiv\theta_4^2/\theta_3^2 $. The set of equations (\ref{th-exp}) may be combined to obtain a power series in $q$.
\begin{eqnarray}
\nonumber
\lambda&\equiv &\frac{1}{2}\left(\frac{\theta_3-\theta_4}{\theta_3+\theta_4}\right)=\frac{1}{2}\left(\frac{1-\sqrt{k'}}{1+\sqrt{k'}}\right) \\
&& =\frac{q+q^9+q^{25}+q^{49}+\cdots}{1+2q^4+2q^{16}+2q^{25}+2q^{36}+\cdots} 
\label{qpower}
\end{eqnarray}
Inverting the series in equation (\ref{qpower}), we obtain 
\begin{equation}
q=\lambda+2\lambda^5+15\lambda^9+150\lambda^{13}+1725\lambda^{17}+ \cdots
\end{equation}

\ack 
This work was supported in part  by  NASA grant NNX08AO04G. We thank Robert Reasenberg (Center for Astrophysics) for his insights and valuable suggestions. We thank Carl Ross at the National Research Council, Canada, for pointing out an error in figure~\ref{fig:3} of an earlier version.

%
\nocite{*}
\end{document}